\documentclass[12pt]{article}
\usepackage[left=1.4in,top=1.25in,right=1.4in,bottom=1.25in]{geometry}
\usepackage{amsmath}
\usepackage{amssymb}
\usepackage{amsbsy}
\usepackage{amsthm}
\usepackage{epsfig}
\usepackage[round]{natbib}
\usepackage{clrscode}
\usepackage{setspace}
\usepackage{multirow}

\newcommand{\N}{\mbox{N}}

\newcommand{\IG}{\mbox{IG}}

\newcommand{\btheta}{\boldsymbol{\theta}}
\newcommand{\bepsilon}{\boldsymbol{\epsilon}}
\newcommand{\by}{\mathbf{y}}
\newcommand{\bx}{\mathbf{x}}
\newcommand{\bz}{\mathbf{z}}
\newcommand{\beff}{\mathbf{f}}

\title{Benchmarking Historical Corporate Performance}
\author{
James G. Scott \footnote{Assistant Professor of Statistics at the McCombs School of Business, University of Texas at Austin.
email: James.Scott@mccombs.utexas.edu.}}
\date{This version: October 2010}

\begin{document}

\maketitle
\begin{abstract}
\noindent This paper uses Bayesian tree models for statistical benchmarking in data sets with awkward marginals and complicated dependence structures.  The method is applied to a very large database on corporate performance over the last four decades.  The results of this study provide a formal basis for making cross-peer-group comparisons among companies in very different industries and operating environments.  This is done by using models for Bayesian multiple hypothesis testing to determine which firms, if any, have systematically outperformed their peer groups over time.  We conclude that systematic outperformance, while it seems to exist, is quite rare worldwide.
\end{abstract}


\section{Introduction}

\subsection{Benchmarking with covariates}

This paper is concerned with the topic of statistical benchmarking---that is, assessing the relative performance of a subject compared to that of its peer group, in a situation where only an absolute measure of performance is available.  Some examples of benchmarking include:
\begin{description}
\item[Mutual-fund ratings,] where the goal is to compare the returns of many funds against their respective benchmarks.  Fund managers who broke even during the financial crisis of 2008, for example, may be superior to those posting double-digit gains during the dot-com bubble of the late 1990's.
\item[University admissions,] where many admissions committees wish to ``adjust'' for a student's economic background and secondary school when comparing absolute measures such as standardized test scores.
\item[Non-overlapping inter-rater comparisons] in subjectively judged competitions like figure skating and high-school debate, where it is natural to control for the fact that different judges may rate participants on highly idiosyncratic scales.
\end{description}

The notion of a peer group is somewhat loose, but can usually be represented through some combination of continuous and categorical measures.  In such cases, it is natural to associate the peer score with the residual from a regression model.  One's peer group can then be interpreted as the group of other subjects having similar expected scores, and one's ``benchmarked score'' as the bit left over after subtracting the expected score, or benchmark.

If a linear model fits the data well, then the residuals from this model are enough to settle the issue.  Complications arise, however, when the marginal distribution of performance is heavy-tailed, highly skewed, or both.  These factors can invalidate traditional parametric assumptions about the distribution of residuals.  It may also simply be very difficult to recover the right ``raw'' marginal distribution through some probabilitistic model that depends upon the covariates.

A further complicating factor is the presence of conditional heteroskedasticity---that is, when the \emph{volatility} of performance differs by peer group, and not merely the expectation.  Here, a benchmark should account for more than simply the absolute value of the residual; it should also adjust for the scale of the residual distribution, given the covariates.  The intuition here is that performing ``further out in the tails," rather than simply differing from the benchmark by a large absolute margin, is the truly impressive---or ignominious---feat.

\subsection{Data, goals, and method}

In this paper, the benchmarking problem is considered in the context of a large data set on historical corporate performance.  Our goal is to use the proposed method to compare publicly traded companies against their peers using a standard accounting measure known as Return on Assets (ROA), which measures the efficiency with which a company's assets are used in generating earnings.  (This is fundamentally different from a market-based measure like stock returns, and is much more stable across time.)  All three complicating factors---skewness, heavy tails, and conditional heteroskedasticity---are present here.  The data set, moreover, is quite large: we have 645,456 records from 53,038 companies in 93 different countries, spanning the years 1966--2008.  This poses serious computational challenges, but also an opportunity for nuanced modeling.

It is necessary to benchmark raw performance numbers because certain features of a company---its industry, its size, its leverage, its country of operation, its market share---make it intrinsically easier or harder to score well on our ROA measure.  These same facts may also entail different levels of volatility in performance.  Such differences, moreover, may be completely unrelated to the differences in managerial talent or performance that we are hoping to measure.  The pharmaceutical industry, for example, is characterized by very high barriers to entry and patent-protected monopolies on specific drugs, allowing these firms to enjoy unusually high returns. 

The proposed benchmarking method is based on two complementary ideas.
\begin{enumerate}
\item First, we account for dependence and heteroskedasticity using flexible Bayesian models applied to an underlying normal latent variable.  We use regression-tree models to account for the highly nonlinear, conditionally heteroskedastic relationships that are present in the corporate performance data.  These ideas have connections with Bayesian models based on Gaussian copulas \citep{hoff:2007a,gramacy:silva:2009}.
\item The results of such a benchmarking analysis are then used to flag firms whose historical performance trajectories indicate that they have systematically outperformed their peers over a period of many years.  To test for the presence of nonzero trajectories, we use Gaussian process priors and construct a Bayesian procedure for large-scale simultaneous testing of all the benchmarked trajectories.
\end{enumerate}

\subsection{Overview of the benchmarking method}

Suppose we observe a response $r_i$, along with a basket of covariates $\bx_i = \{x_{ij} \}_{j=1}^p$, for each of $N$ subjects.  Suppose further that the $r_i$ arise marginally from a population whose cumulative distribution function is $F$.  Then clearly $r_i = F^{-1}(u_i)$, where $F^{-1}$ is the pseudo-inverse of the CDF, and where $u_i$ is uniformly distributed on $(0,1)$ marginally.  One may in turn express $u_i$ in terms of a standard normal variable $z_i$ using the probit link:
$$
r_i = F^{-1}\{\Phi(z_i)\} \, ,
$$
where $\Phi$ is the standard-normal CDF---in essence, a univariate Gaussian copula model.

Given $F$, it is possible to model the conditional distribution of $z_i$, given $\bx_i$, and to assess how far out in the tails the ``benchmarked'' value of $z_i$ lies.  For example, a simple method would be to represent each latent variable as $z_i = \mu_i + \sigma_i y_i$.  The benchmark score for firm $i$ can then be associated with the standardized residual $y_i$,
$$
y_i = \frac{z_i - \mu_i}{\sigma_i} \, .
$$
This quantity clearly adjusts both for the expected performance and the expected volatility of performance, given $\bx_i$.  The idiosyncratic means and scales are entirely consistent with the notion that the $z_i$'s are, take as a whole, standard normal.

In this simple example, the expression for $z_i$ is overspecified without imposing structural relationships among the means and variances.  It also depends upon knowing the distribution function for $y$.  Hence we need the following:
\begin{enumerate}
\item A method for handling the marginal distribution $F$, which is in some sense a nuisance parameter.
\item A model for the conditional expectation $\mu_i$, given the covariates $\bx_i$.
\item A model for the conditional variance $\sigma^2_i$, given the covariates $\bx_i$.
\end{enumerate}

In the corporate performance database, the measurements are continuous and $N$ is very large.  It is therefore reasonable to estimate $F$ using the empirical distribution function.  Denote this estimate by $\hat{F}$, and define $\hat{z}_i$ as
$$
\hat{z}_i = \Phi^{-1} \left\{ \frac{N}{N+1} \hat{F}(  r_i) \right\} \, ,
$$
rescaling slightly to avoid evaluating $\Phi^{-1}$ exactly at 1.   The Dvoretzky--Kiefer--Wolfowitz inequality ensures that, since $N$ is large, $F$ and $\hat{F}$ will be close in $\sup$ norm.  Under this empirical-Bayes approach, one can directly specify a model for the conditional mean and variance of the transformed data $\hat{\bz} = (\hat{z}_1, \ldots, \hat{z}_N)^t$, making full use of normality assumptions that are not even approximately satisfied by the original data.  (The method would be essentially unchanged if $\hat{F}$ were computed using some other procedure.)

In other situations, however, $N$ may be too small for $\hat{F}$ to be a stable estimator of the CDF, or the data may be ordinal (in which case $F^{-1}\{\Phi(\cdot) \}$ is a many-to-one function).  In these cases, two good options are available.  Both can be implemented by constructing a Markov chain having a specified stationary distribution for the latent variables $z_i$, which allow one to compute erdogic averages for the benchmark scores $y_i$.

In the first option, this stationary distribution is the full Bayesian posterior for $\bz$, given the assumed model.  In the second, the stationary distribution is a pseudo-posterior that results from a particular simplification to the marginal likelihood due to \citet{hoff:2007a}.  Additionally, the second option opens up the possibility of a direct Monte Carlo sampler from the pseudo-posterior, avoiding the potential difficulties associated with MCMC in large dimensions.  Neither of these options are considered here, but may be of interest to those interesting in benchmarking for data sets where $N$ is more moderate.

\section{Background and exploratory analysis}

Understanding the reasons why some firms thrive and others fail is one of the primary goals of research in strategic management.  Studies that examine successful companies to uncover some ``secret recipe'' for success are very popular, both in the academic and popular literature.

Before the search for special causes can begin, however, success must be quantified and benchmarked.  In this paper, we use a common metric called ``return on assets'' (ROA), which gives investors some notion of how effectively a company uses its available funds to produce income. Higher ROA numbers are better, since they imply that a company is doing more with less.

But comparing ROA numbers across different industries, different eras, or different countries poses special challenges.  We must, in effect, transform all data to a common scale in order to facilitate an apples-to-apples comparison of companies in different peer groups.

Much academic work in this direction focuses on decomposing observed variation in corporate outcomes into sources that arise from market, industry, and firm-level processes \citep{bowman:helfat:2001, hawawini:etal:2003}.  Other work is built on clustering algorithms \citep[for example,][]{harrigan:1985} that attempt to find logical performance-based groupings of firms.

The goal of this paper is different: we wish to adjust for the effects of covariates to produce a pure performance measure, one that is largely independent of factors beyond the control of a firm's managers and employees.  We wish to do this, moreover, for a very large database: as was mentioned in the introduction, we have 645,456 annual ROA observations on 53,038 companies in 93 countries, spanning 1966--2008.  This encompasses over $85\%$ of all publicly traded companies in all industrialized countries for the last four decades.  Having done this, we seek to identify companies with sustained patterns of non-random performance over time \citep{ruefli:wiggins:2000,Scott:2008}.

\begin{figure}
\begin{center}
\includegraphics[width=4.5in]{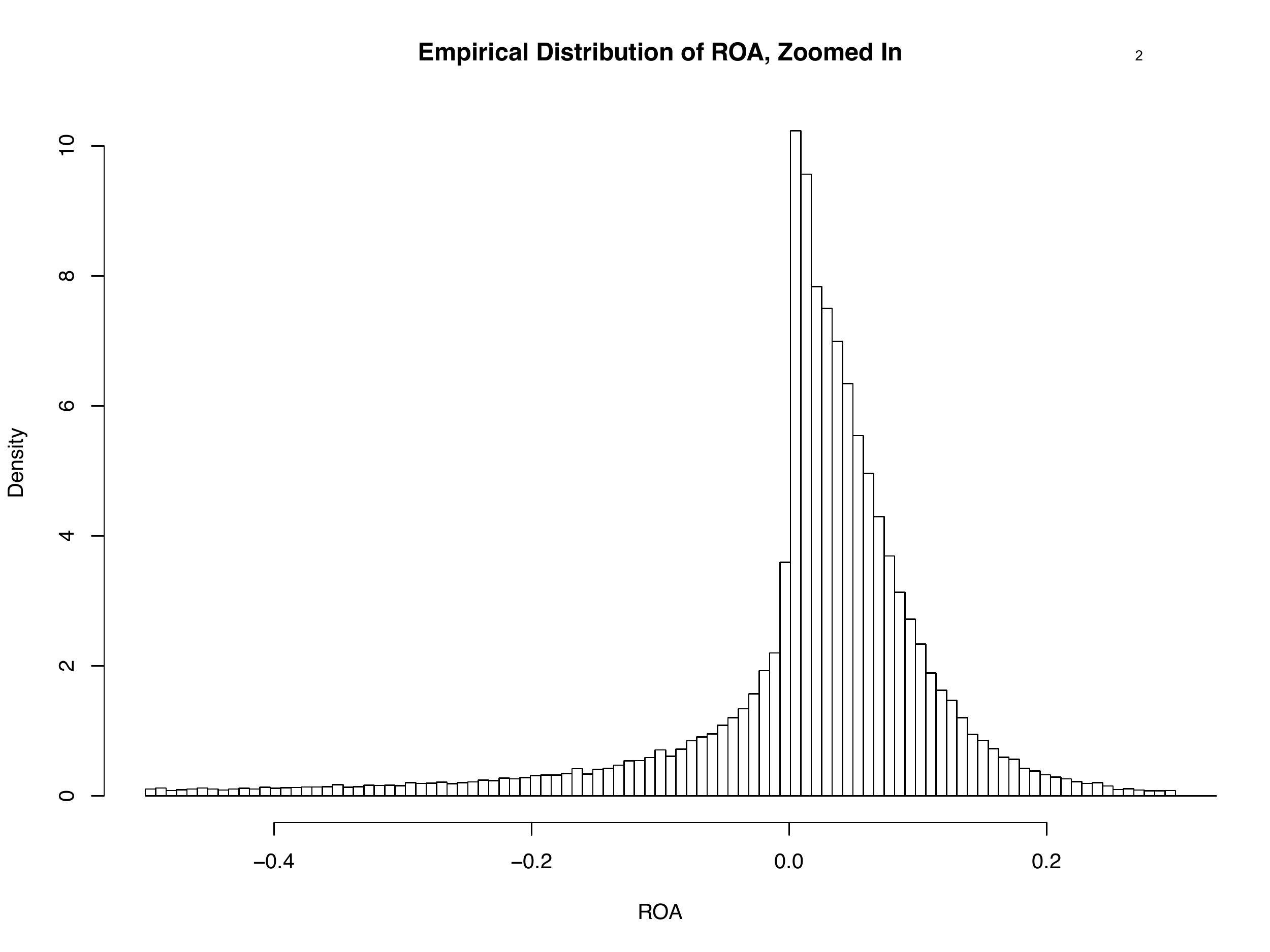}
\caption{\label{ROA-raw-figure} The empirical distribution of return on assets (ROA) in the cohort under study.}
\end{center}
\end{figure}

This database, however, exhibits a number of features that make statistical benchmarking difficult.  Some of these features were mentioned in the introduction, but to recap the main ones:
\begin{description}
\item[Skewness and heavy tails:] The empirical distribution of observed ROA values has a heavier left tail (i.e.~negative values) than right tail (positive values).  Evidently it is far more common to perform terribly than to perform excellently.  Figure \ref{ROA-raw-figure} shows the worldwide distribution of ROA, zoomed in to show the range from $-50\%$ to $+30\%$.  Notice the extremely long lower tail (which extends far beyond the window shown here); the peakedness near zero; and the concentration of mass between $0\%$ and $10\%$. 
\item[Heteroskedasticity:] Information such as industry, size, debt, and market share seem to be relevant in predicting both the expected value (center) and expected spread (volatility) in observed ROA values.
\item[Missingness and sample-size heterogeneity:] Some companies have as little as 1 year of data; other companies have as many as 41.  Similarly, there are gaps in many of the observed time series---for example, we observe a company from 1980--1985 and 1988--2000, but not in 1986--1987.
\end{description}

\begin{figure}
\begin{center}
\includegraphics[width= 4.5in]{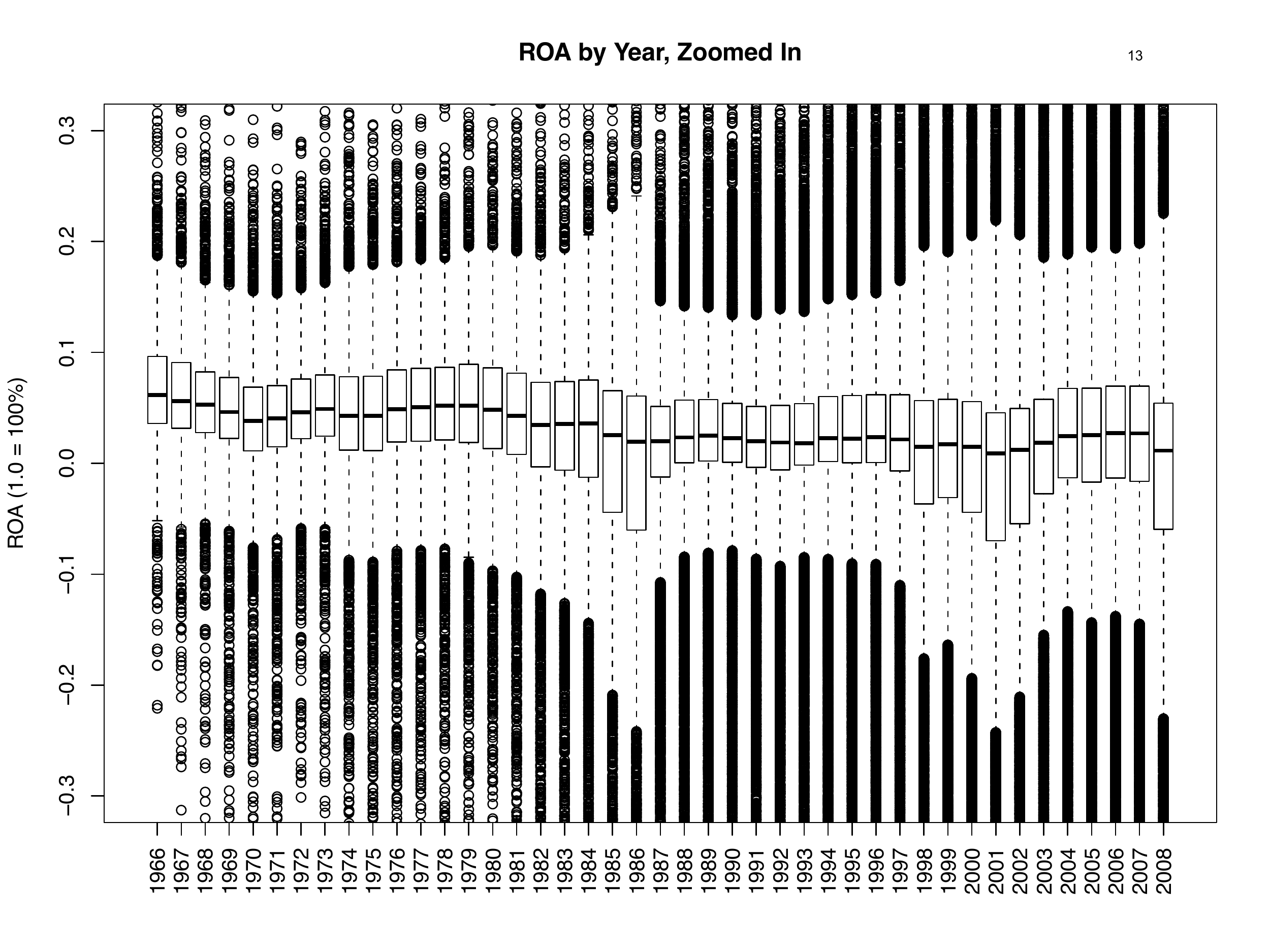}
\caption{\label{ROA-EDA1-figure} Boxplots showing large differences in the distribution of raw ROA across different years.}
\end{center}
\end{figure}

\begin{figure}
\begin{center}
\includegraphics[width= 4.5in]{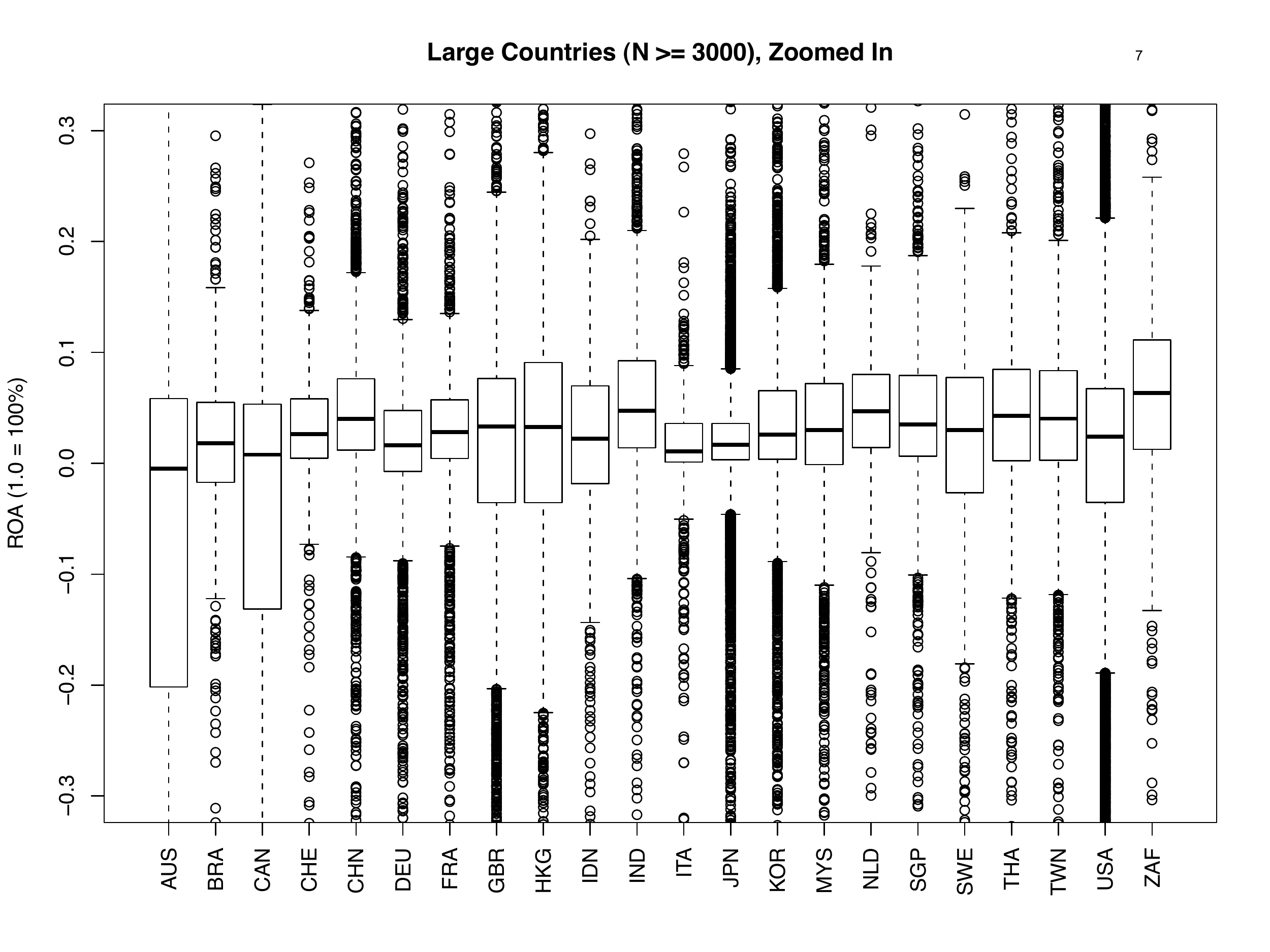}
\caption{\label{ROA-EDA2-figure} Boxplots showing large differences in the distribution of raw ROA across different countries.  (Only the 22 largest countries by sample size are included.)}
\end{center}
\end{figure}

\begin{figure}
\begin{center}
\includegraphics[width= 4.5in]{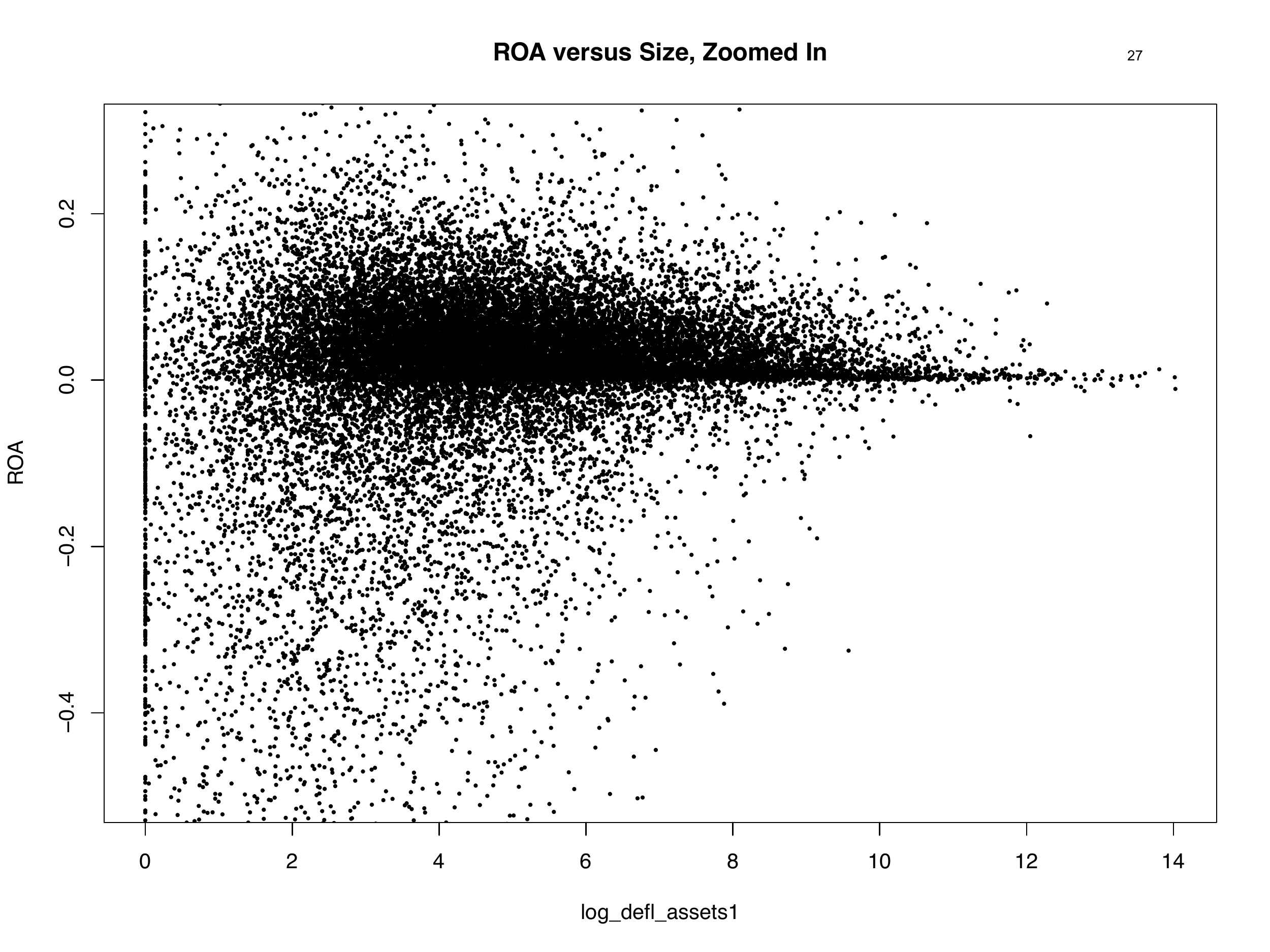}
\caption{\label{ROA-EDA3-figure} ROA versus inflation-adjusted assets.}
\end{center}
\end{figure}

\begin{figure}
\begin{center}
\includegraphics[width= 4.5in]{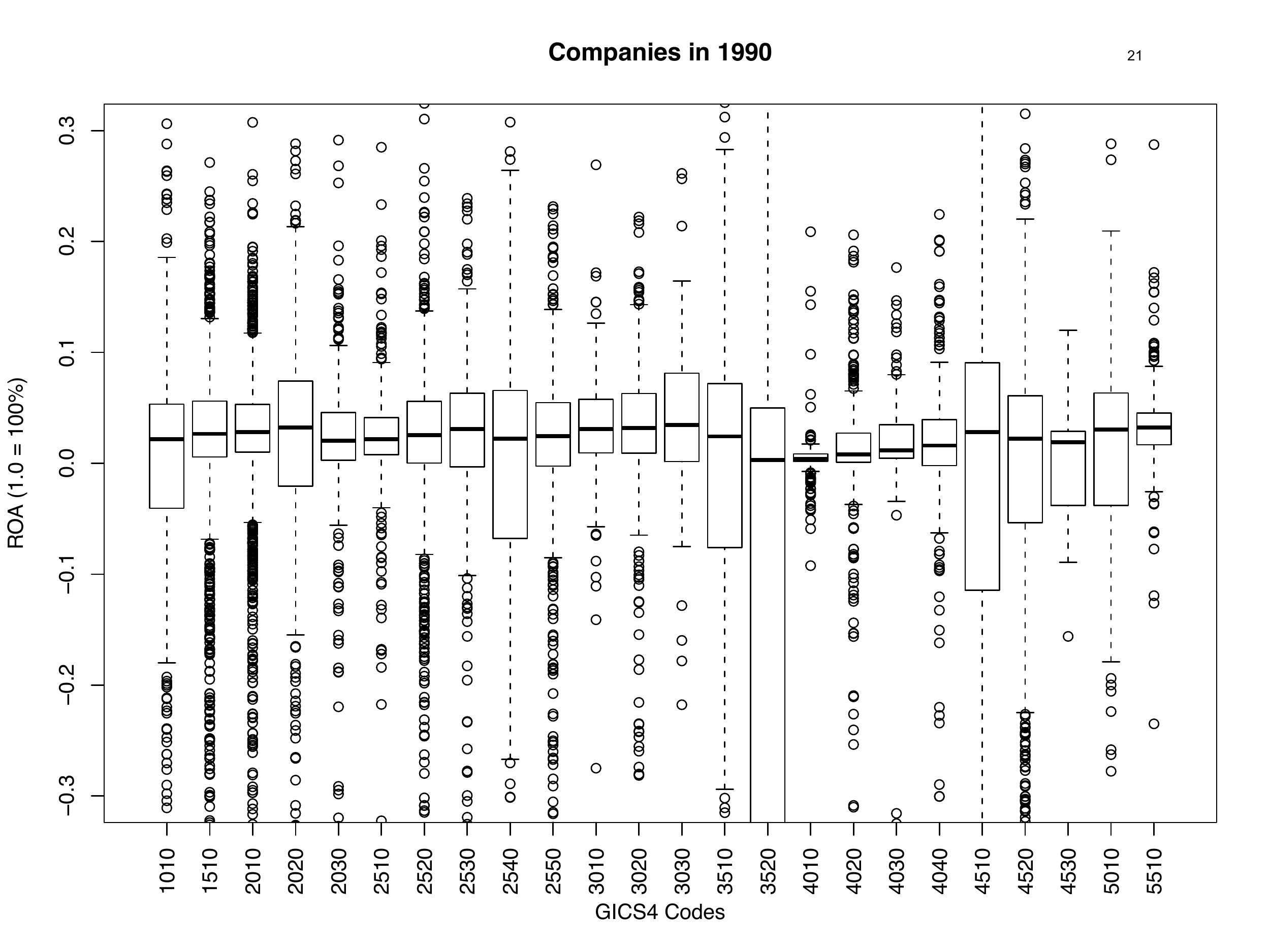}
\caption{\label{ROA-EDA4-figure} ROA in 1990, stratified by the groupings in the four-digit Global Industry Classification System (GICS).}
\end{center}
\end{figure}

To get a sense of how ROA depends upon the relevant covariates, consult Figures \ref{ROA-EDA1-figure}--\ref{ROA-EDA4-figure}.

Figure \ref{ROA-EDA1-figure} plots the marginal distribution of ROA as it changes over time.  Clearly there are large temporal fluctuations caused by major macroeconomic trends.

Figure \ref{ROA-EDA2-figure} shows the ROA distributions for the 22 countries in the database with the most number of records.  Major cross-country differences are apparent, both in the center and spread of the ROA distribution.  Australia, for example, has an enormous left tail compared to other countries; Britain, Hong Kong, and the United States appear structurally similar to each other, though with some significant differences; China and India have very small lower tails and a much larger concentration of companies with positive ROA; and Japan and Italy are characterized by upper and lower tails that are much thinner than those of other countries in the sample.

Figure \ref{ROA-EDA3-figure}, meanwhile, shows ROA versus size, as measured by the log of a company's inflation-adjusted assets.  There is a clear heteroskedastic ``fan'' shape that persists over time: low-asset companies tend to have more variable ROA than high-asset companies.  High-asset companies also tend to have higher returns than low--asset companies, though this tendency is not as strong.

Figure \ref{ROA-EDA4-figure} shows a snapshot of global ROA versus industry for 1990.  The industry groupings are taken from the Global Industry Classification System, or GICS.  One can see large systematic differences due to industry group.  For example:
\begin{itemize}
\item Biotechnology and pharmaceutical firms (3520) and information-technology firms (4510, 4520, and 4530) tend to have ROA distributions that are very spread out.
\item Commercial banks and mortgage companies (4010) have an ROA distribution that is very narrowly peaked around zero, with occasional outliers.
\end{itemize}
We have shown only the boxplots for 1990, and while these differences are relatively stable, some aspects of the relationship between industry group and ROA do change over time.  Some of these differences can be explained by different levels of assets in each industry group.  Even after adjusting for assets, however, some systematic differences in the center and spread of the ROA distribution remain.  It is clear from these explorations that we must build a model that allows potential interactions of industry group and year in order to account for observed peer-group effects, including the obvious conditional heteroskedasticity apparent in the plots.

\section{Benchmarking with tree models}

\subsection{Bayesian regression trees}

To account for all these nuances in the data, we fit a Bayesian regression-tree model to the transformed Gaussian $z_i$'s corresponding to the raw ROA values.

Bayesian tree models are described in detail by \citet{denison:mallick:smith:1998}, \citet{chipman:george:mcculloch:1998}, and \citet{gramacy:lee:2008}, building upon the classical work by \citet{breiman:friedman:etal:1984}.  The goal of the method is to use a tree with binary splitting rules involving the covariates to model the conditional distribution of $z_i$.  A tree itself is best understood as a partition of the data points into a mutually exclusive, collectively exhaustive set of ``terminal nodes.''  Each data point $(z_i, \bx_i)$ is assigned to a terminal node of the tree by ``cascading'' down the tree, and successively applying the binary splitting rules at each non-terminal node to determine whether the data point goes into the left daughter node or the right daughter node.  Each splitting rule is a binary decision involving one of the covariates $x_j$, and the`` cascade'' is guaranteed to stop, since every legal tree has a finite maximum depth.

The prior over tree space is defined implicitly by a process for generating random trees.  When combined with the likelihood for $z_i$'s in each terminal node, this produces a posterior distribution over tree space, and therefore a posterior over the supported family of conditional distributions of $z_i$, given the covariates.  This prior has two parts: a prior over the topology or shape of the tree, and a prior over splitting rules that define each non-terminal node of the tree.

The prior over tree topology is indexed by two parameters $\alpha$ and $\beta$, which describe the probability that a given node $K_t$ will be split into two further nodes:
$$
p(\mbox{split }K_t) = \alpha \{1+d(K_t)\}^{-\beta} \, 
$$
where $d(K_t)$ is the depth of the node $K_t$ (i.e.~how many previous splits have take place to yield $K_t$ as a terminal node).  As a default choice, we choose $\alpha = 1/2$ and $\beta = 2$.  This has the effect of regularizing the conditional probability model for $z_i$ by limiting the size of the tree, thereby preventing overfitting.

Finally, the prior over splitting rules is defined hierarchically, as follows.  For a given node, the covariate used in the split is chosen uniformly from among the $p$ covariates under consideration.  Then once a particular covariate has been chosen, the value that defines the split region is chosen uniformly from the empirical distribution of observed covariate values in the data set.

Trees partition the predictor space into hypercubes, represented by terminal nodes on the tree, such that the distribution of outcomes in each hypercube is relatively homogenous.  Given the assumed normality of the $z_i$'s, it is reasonable to describe the data distribution in each terminal node $K_t \in \mathcal{K}$ using a normal distribution with mean $\mu_{K_t}$ and standard deviation $\sigma_{K_t}$. This assumption allows the marginal likelihood of a proposed tree to be calculated in closed form.  To fit this model, the Metropolis--Hastings algorithm is used to explore the space of trees.  The method is explained in detail by \citet{chipman:george:mcculloch:1998} and \citet{gramacy:lee:2008}, and is implemented in the R package \verb|tgp| \citep{Gramacy:Taddy:2009}.

\subsection{Results of model fit}

\begin{figure}
\begin{center}
\includegraphics[width=5.5in]{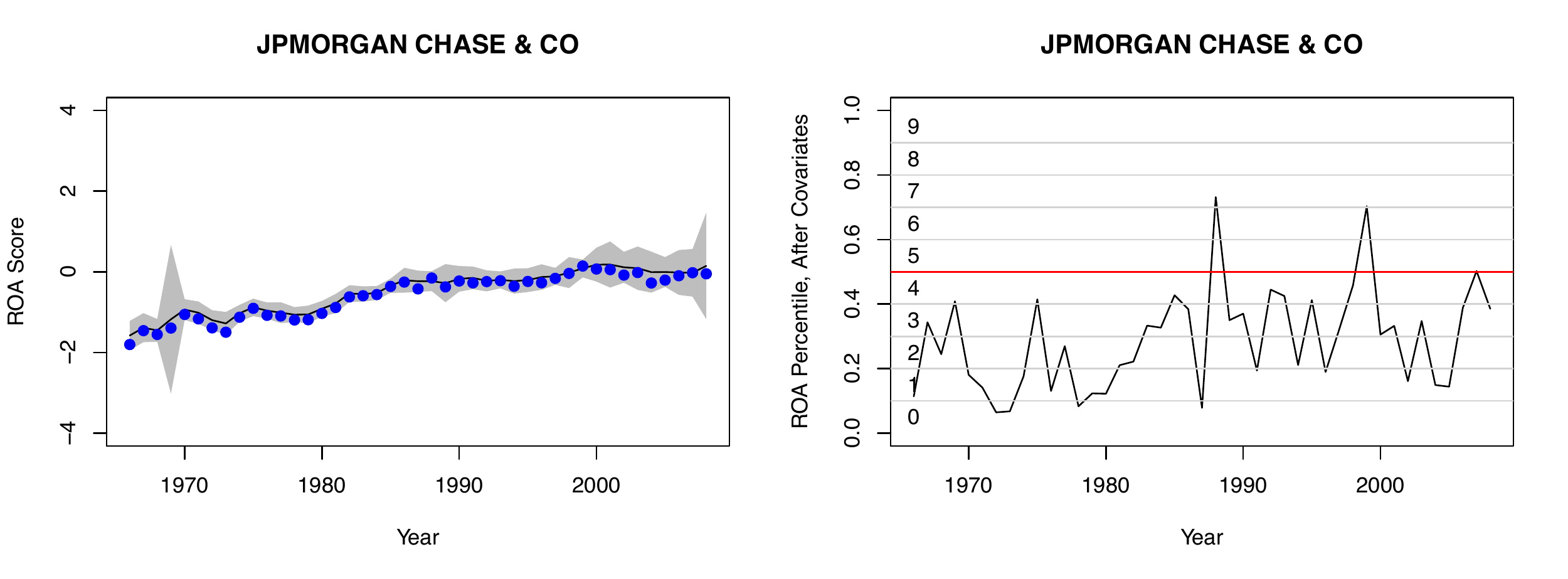}
\includegraphics[width=5.5in]{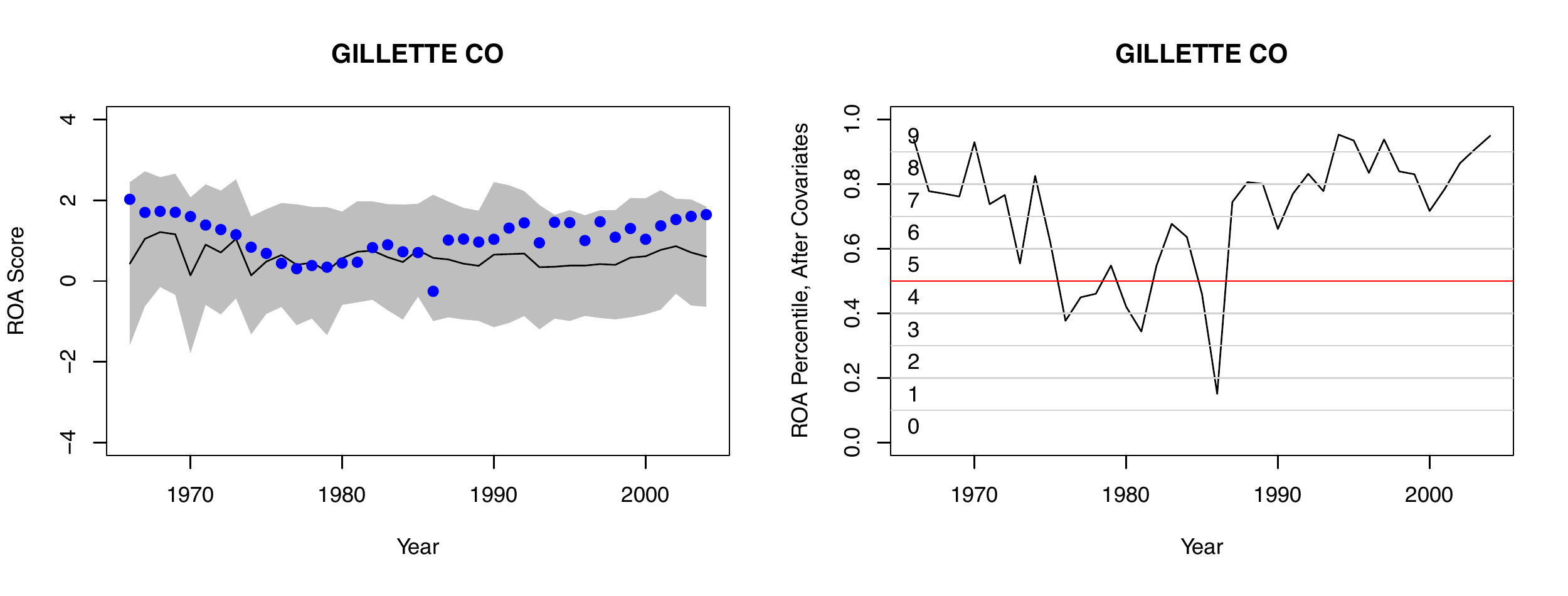}
\includegraphics[width=5.5in]{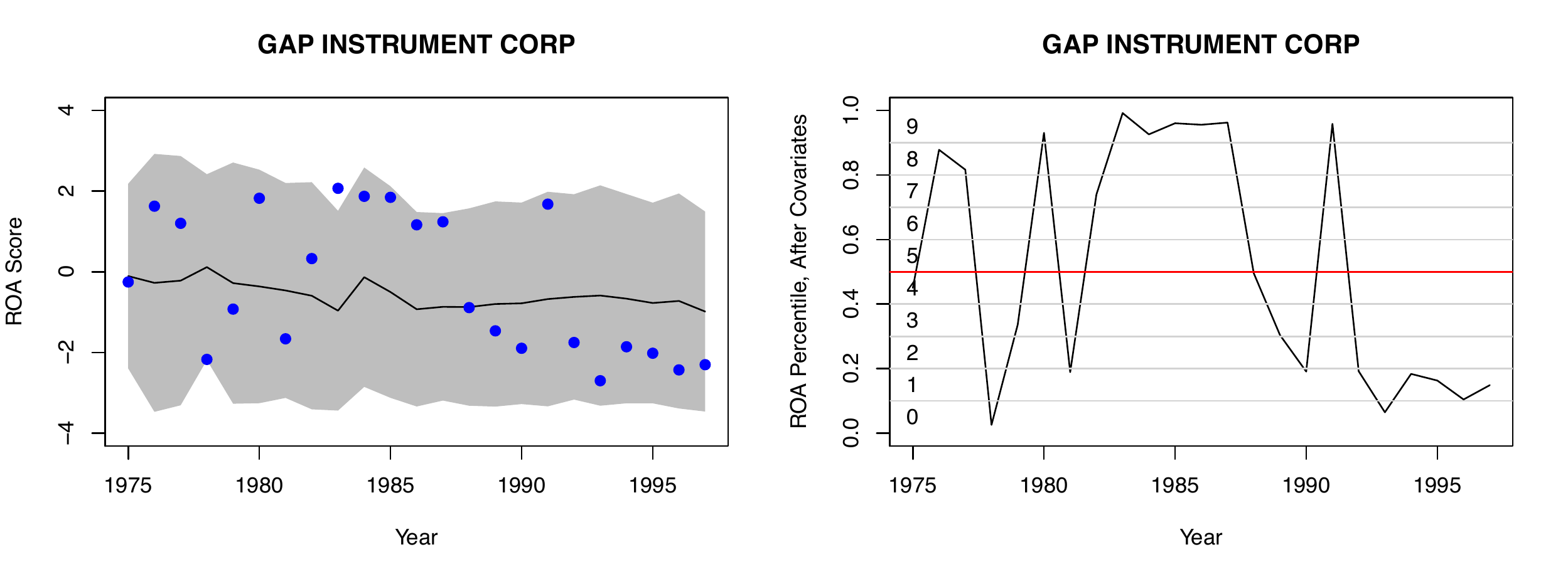}
\caption{\label{fig:examples1}Before and after pictures of some example companies.  The left column shows the benchmark (black line); conditional $95\%$ predictive interval for $z_{it}$, given $\bx_{ij}$ (shaded area); and actual value of $z_{ij}$ (dots).  The right column shows the same companies placed onto a common $(0,1)$ scale of benchmarked performance, $\Phi\{(z_{ij} - \mu_{ij}) / \sigma_{ij} \}$.  }
\end{center}
\end{figure}

The tree model described in the previous subsection was used to fit the conditional distribution of $z_i$, given covariates $\bx_i$.  We use the following covariates to propose splitting rules for the tree:
\begin{description}
\item[Year:] The year of the observed ROA value, to account for differences over time.
\item[GICS:] A variable indicating a company's code in the GICS classification system, a standard method for classifying firms into successively finer, more homogenous groups.  GICS codes are 8 digits and rigidly hierarchical.  For example, all companies whose codes begin with 25 are in the ``Consumer Discretionary'' economic sector.  This is, in turn broken down into further categories by the next three pairs digits of the code: e.g.~2550 for ``Retail,'' 255040 for ``Specialty Retail,'' and 25504010 for ``Apparel Retail.''  This hierarchical structure is mirrored across all categories.  To capture it, we simply treat the GICS code as a numerical variable to be used directly in tree splitting rules.
\item[Log deflated assets,] a proxy for company size that adjusts for differences due to the effects of inflation over the last four decades.
\item[Leverage,] which captures a company's ratio of debt to assets.  Highly leveraged firms tend to have more volatile performance patterns, meaning that this covariate will be especially useful for capturing differences in the expected magnitude departure of $z_i$ from its conditional mean.
\item[Country of operation:] to use this as a covariate, we computed the pairwise Kolmogorov-Smirnov (KS) distance between the empirical distribution of ROA in the United States, and that of every other country in the sample.  All firms from country X were assigned a covariate value defined by this USA--X pairwise KS distance.  Hence countries whose empirical ROA distributions are more similar are more likely to be near each other in tree space.  In principle, any country could have been chosen as a baseline for this KS-distance calculation, but the USA was used because it had the largest number of samples.
\end{description}
It is not necessary to include interaction terms explicitly, since these are handled automatically by the rich structure of the tree model.

Given a particular tree, we can easily compute benchmarked residuals.  Let $\gamma_i^{(T)}$ denote the tree node associated with observation $z_i$ at step $T$ of the Monte Carlo sample.  The conditional benchmark score can then easily be calculated as:
$$
y_i^{(T)} = \frac{z_i - \mu_{\gamma_i^{(T)}}}{\sigma_{\gamma_i^{(T)}}} \, ,
$$
given the tree parameters and node allocations.  We then define $y_i$, the estimated posterior mean for the benchmark score, as the benchmark score averaging over posterior uncertainty about the true tree model.  This is well approximated by the ergodic average of $y_i^{(T)}$ over the MCMC samples from tree space.

It is not possible to show the output for all 53,038 companies, but Figure \ref{fig:examples1} shows three examples: JPMorgan, a large financial-services firm; Gillette, which manufactures skin-care products; and Gap Instrument Corporation, a small software firm.  The figure gives a ``before and after'' snapshot for the three firms, demonstrating how the tree model effectively adjusts for both the expected value, and the expected volatility, of performance.  The left column shows the benchmark mean (black line); the conditional $95\%$ predictive interval for $z_{it}$, given $\bx_{ij}$ (shaded grey area); and actual value of $z_{ij}$ (dots).  Notice how the benchmark distribution for the small software firm (Gap) is much wider than the benchmark distribution for the large investment bank (JPMorgan).  This reflects the fact that, empirically, the performance of small firms is much more variable during the period studied than the performance of the investment banks.

The case of JPMorgan is especially interesting, because one can see the benchmark distribution widen during two notable time periods: the recession of 1969--70, and the financial panic of 2008.  This reflects the relatively higher degree of volatility experience by financial firms during these two episodes.

The right column of Figure \ref{fig:examples1} then shows the same three companies placed onto a common $(0,1)$ scale of benchmarked performance, defined by the CDF of a standard normal distribution: $\Phi(y_{ij}) = \Phi\{(z_{ij} - \mu_{ij}) / \sigma_{ij} \}$.  These plots, which can be generated for all 53,038 firms, show how the Bayesian tree model effectively benchmarks firms against their peer groups, facilitating fair comparisons across a wide variety of industries and operating environments.

\section{Simultaneous testing of benchmarked data}

\subsection{Notation and general approach}

Let $z_{it}$ denote the benchmarked observation for company $i$ at time $t$; let $\mathbf{z}_i$ denote the whole vector of benchmarked observations for company $i$ observed at times $\mathbf{t}_i = (t_1, \ldots, t_{n_i})$; and let, let $Z$ denote the set of $\bz_i$ for all $i$.  Our general approach is to decompose each observed trajectory $\by_i$ as $\bz_i = \beff_i + \bepsilon_i$, where $\beff_i$ represents an unobserved mean trajectory and $\bepsilon_i$ a vector of noise.

By ``noise,'' we do not mean measurement error, but rather short-term variations in performance that are not suggestive of any longer-term trends.  This model specification reflects the growing recognition in the management-theory literature that one must account for potential autocorrelation in performance over time.  See, for example, \citet{denrell2003}, \citet{denrell2005}, and \citet{henderson:etal:2009}.  This assumption allows for long runs of above- or below-average performance due simply to luck. 

Accordingly, we model each benchmarked trajectory having an autoregressive component.  Let $\btheta = (\phi, v)$, and let $\Sigma_{\btheta}$ denote the familiar AR(1) covariance matrix having diagonal entries $\Sigma_{ii} = v/(1-\phi^2)$ and off-diagonal entries
$$
\Sigma_{ij} = \left(\frac{v}{1-\phi^2} \right) \phi^{|i-j|} \, .
$$

We assume the following model:
\begin{align}
\bz_i &\sim \N(\bz_i \mid \mathbf{f}_i, \Sigma_{\btheta_i}) \label{simple1} \\
\btheta_i &\sim U(\phi_i \in (0,1)) \times \IG(v_i \mid a, b) \label{simple2} \\
\mathbf{f}_i &\sim w \cdot G(\mathbf{f}_i) + (1-w) \delta_{F_0} \label{simple3} \, ,
\end{align}
where $\Sigma_{\theta}$ is the familiar AR(1) variance matrix; $\delta_{F_0}$ is a Dirac measure placing positive probability on the hypothesis that $\beff_i = F_0$, the null function that is zero everywhere; $w \in [0,1]$ is the prior probability of having a nonzero mean trajectory; and $G(\beff_i)$ is distribution over possible nonzero mean trajectories.

Even though the global distribution of benchmarked ROA values will be standard normal, the trajectories of individual firms can still display very different patterns of autocorrelation and marginal variation.  

It is convenient to introduce a model-index parameter $\gamma_i$ which takes the value 0 if $\beff_i = F_0$, and 1 otherwise.  The posterior probabilities $w_i = \mbox{pr}(\gamma_i \neq 0 \mid Y)$ then can be used to flag non-null units.  These are called the posterior inclusion probabilities, reflecting inclusion in the non-null set.

\subsection{Bayesian adjustment for multiplicity}

We have phrased the problem of identifying nonrandom performers as one of large-scale, simultaneous Bayesian testing of benchmarked trajectories, and must therefore manage the multiplicity problem that arises.  Bayesian multiple-testing procedures have been extensively studied in many different contexts.  See \citet{scottberger06} and \citet{bogdan:ghosh:2008b} for theoretical development;  both \citet{domuller2005} and \citet{dahl:newton:2007} exhibit the use of such methods in large genomic problems.  Additionally, \citet{muller:etal:2006}, \citet{bogdan:etal:2008}, and \citet{park:ghosh:2010} go into detail about the relationship between Bayesian multiple testing and classical approaches that control the false-discovery rate.

The focus here is not on developing new methodology for multiple testing, but rather on applying existing methodology to the benchmarked trajectories from the corporate-performance database.  These methods have the attractive property than they can adjust for multiplicity automatically, without the need for using ad-hoc penalty terms to control the false-discovery rate.

The model must be completed by specifying a model for nonzero trajectories, along with priors for $\phi_i$ and $v_i$.  We place a uniform prior on $\phi_i \in (0,1)$, along with a conjugate Inverse-Gamma$(5/2,5/2)$ prior on $v_i$.  To model the nonzero trajectories, we use a Gaussian process prior with a squared-exponential covariance function.  This ensures that the resulting trajectories will be smooth, i.e.~have derivatives of all order.  The Gaussian-process model is described in detail by \citet{Scott:2008}, wherein a similar method is applied to a much smaller database of 3,459 firms, and where an MCMC sampling algorithm is described for computing posterior inclusion probabilities.

The model solves both of the major problems encountered in the ROA data: time-varying nonzero trajectories, and the possibility of long-run departures from the average due to the autoregressive behavior of the residuals.

\subsection{Results}

\begin{figure}
\begin{center}
\includegraphics[width=5.5in]{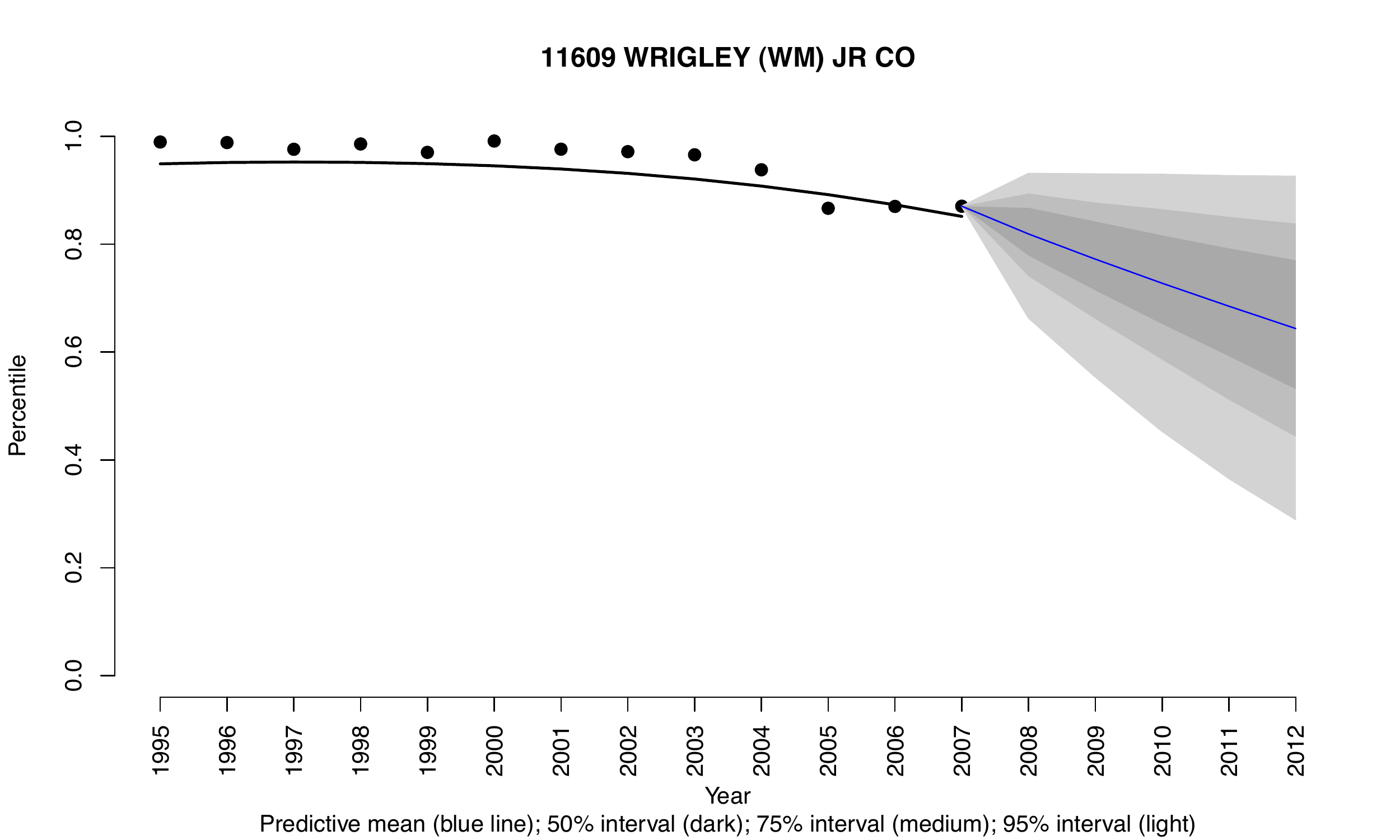}\\
\includegraphics[width=5.5in]{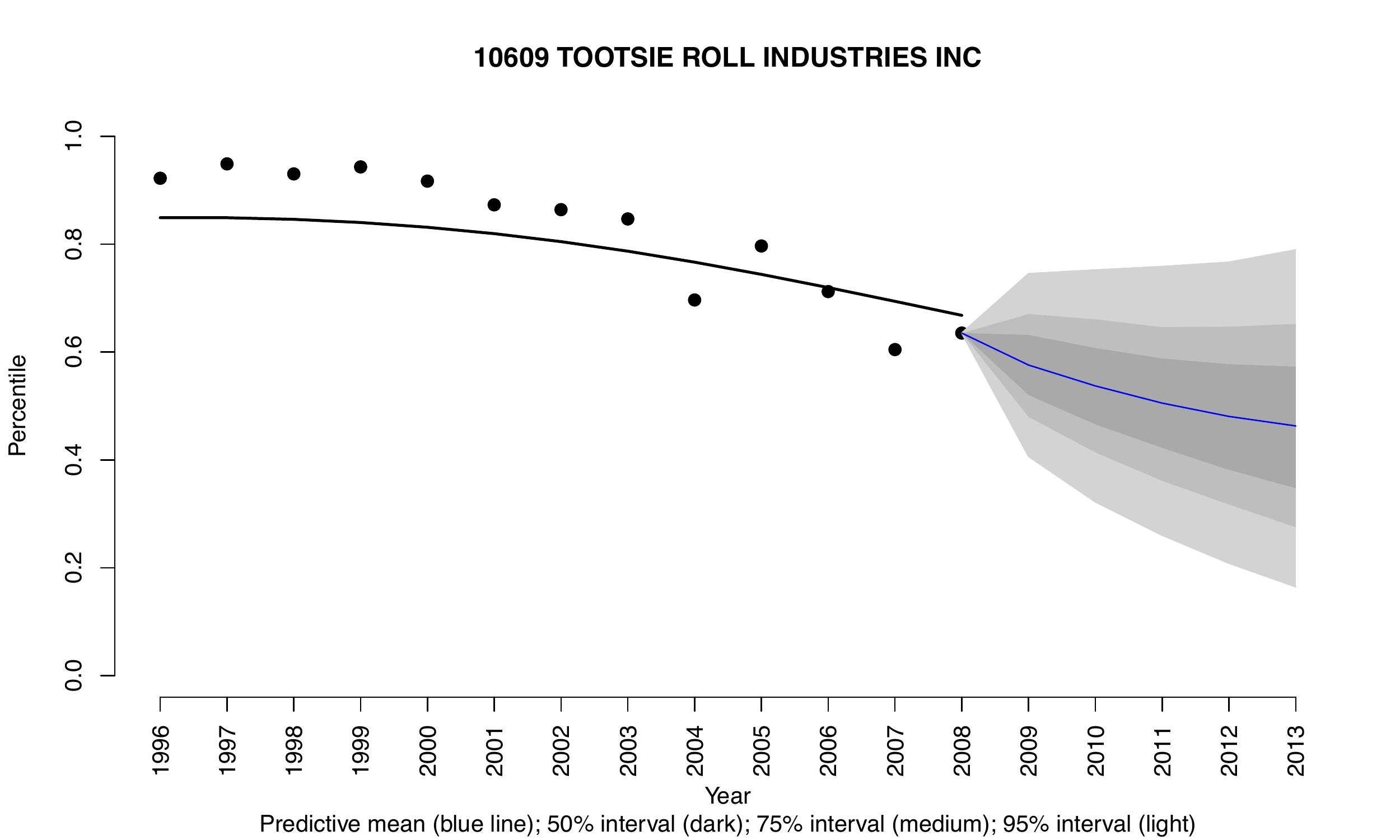}
\caption{\label{fig:examples2}Comparison of two manufacturers of sweets, Wrigley and Tootsie Roll.  Wrigley was flagged by the multiple-testing procedure as out-performing its peer group, while Tootsie Roll was not.}
\end{center}
\end{figure}

\begin{figure}
\begin{center}
\includegraphics[width=5.5in]{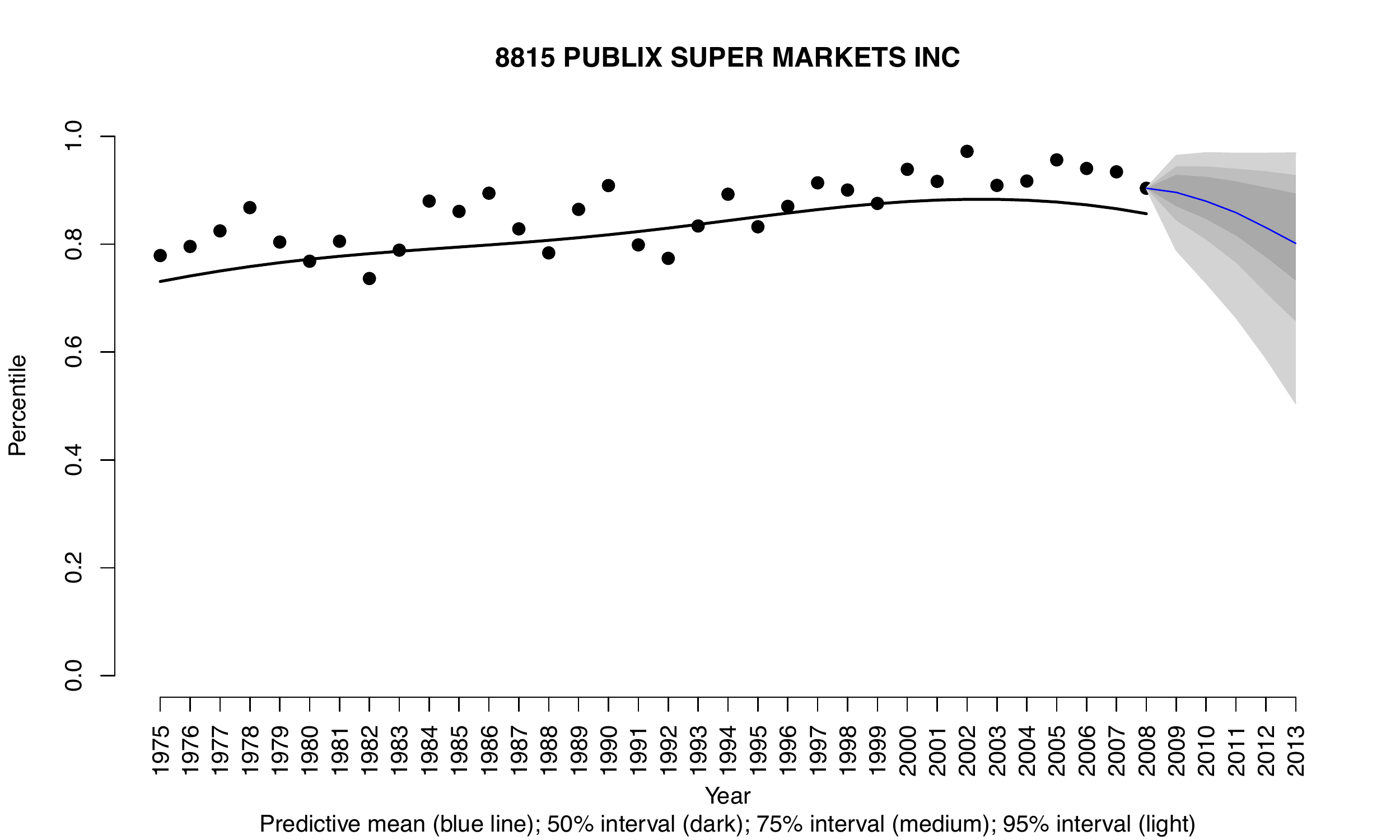}\\
\includegraphics[width=5.5in]{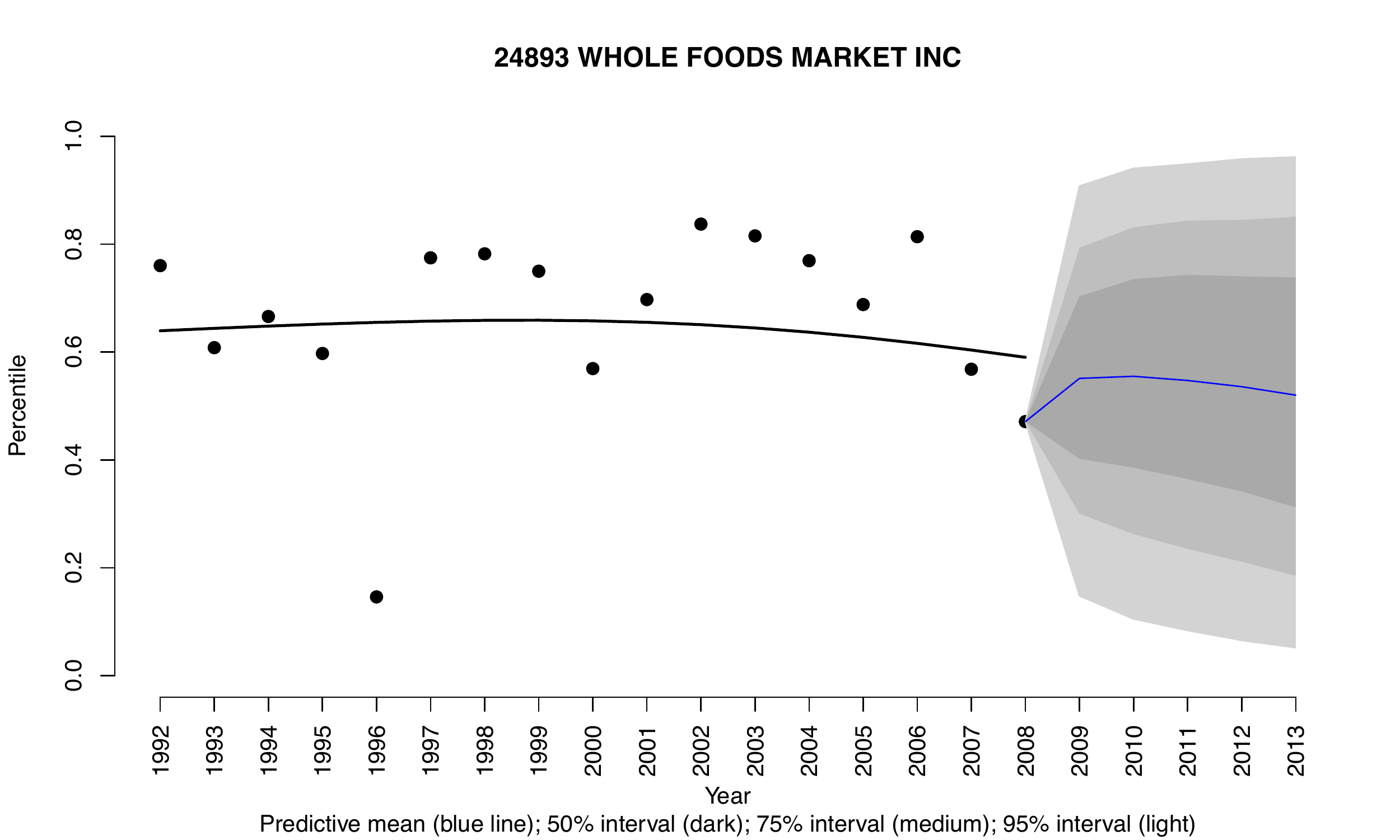}
\caption{\label{fig:examples3}Comparison of two grocery store chains, Publix and Whole Foods.  Publix was flagged by the multiple-testing procedure as out-performing its peer group, while Whole Foods was not.}
\end{center}
\end{figure}

\begin{figure}
\begin{center}
\includegraphics[width=5.5in]{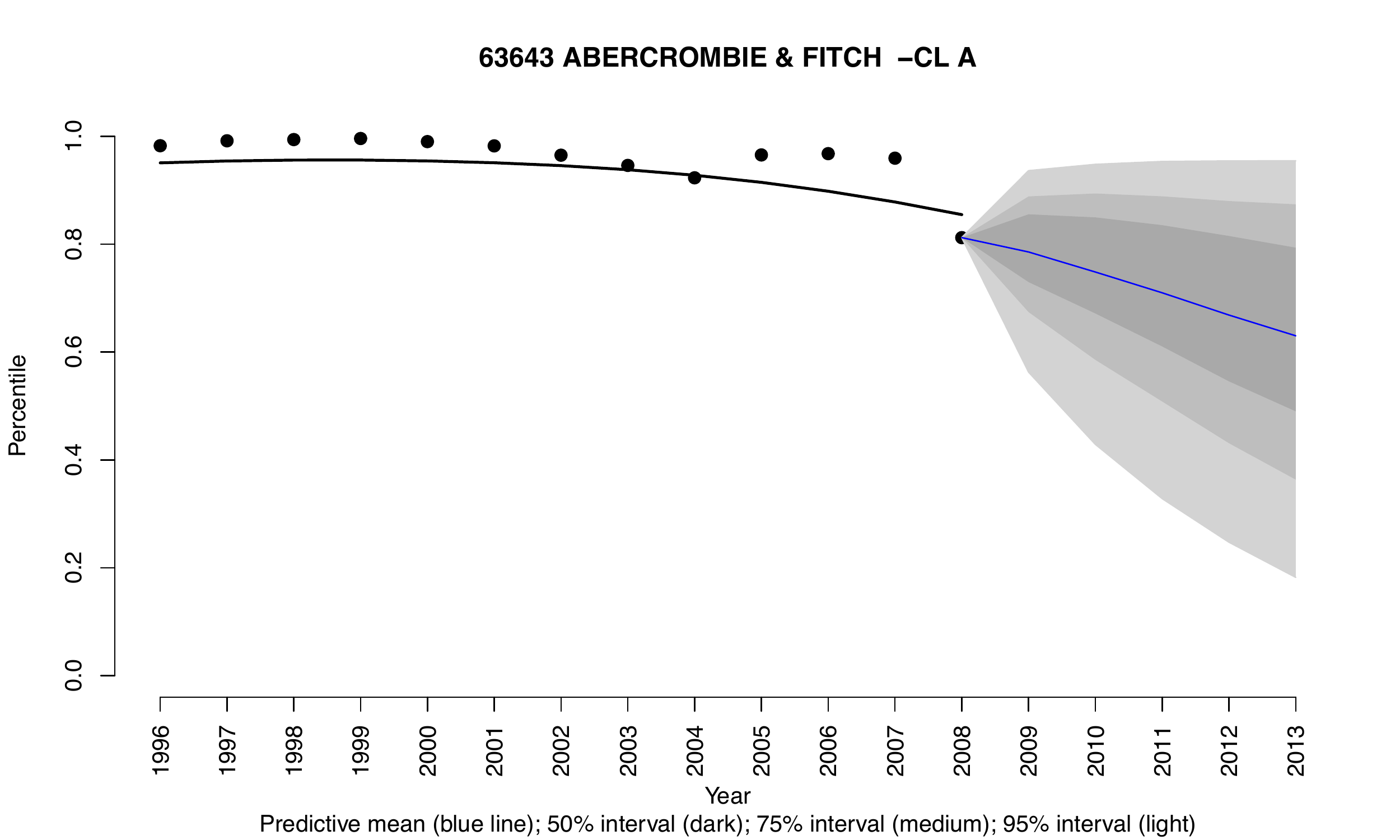}\\
\includegraphics[width=5.5in]{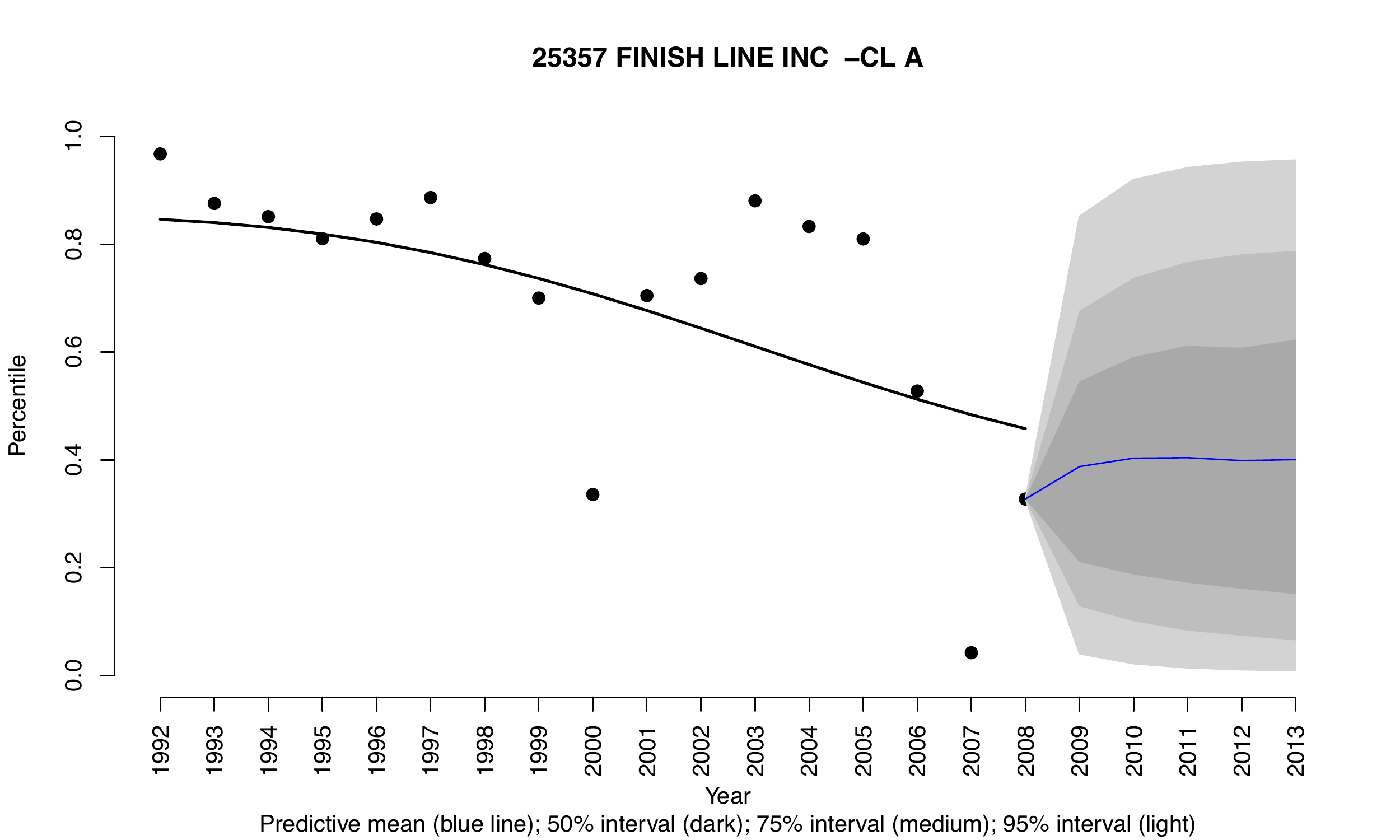}
\caption{\label{fig:examples4}Comparison of two apparel companies, Abercrombie and Finish Line Sports.  Abercrombie was flagged by the multiple-testing procedure as out-performing its peer group, while Finish Line was not.}
\end{center}
\end{figure}

Given the nature of the data, many interesting summaries of the results are available.  We choose to focus on three examples, which give the reader a notion of the use to which these results might be put.  Each example is a comparison of two firms in the same general area of business: two candy-making companies, two grocery-store chains, and two apparel manufacturers.  In each pair, one firm was flagged by the Bayesian multiple-testing procedure as having systematically outperformed its peer group with probability greater than $95\%$.  The other member of the pair had less than a $50\%$ probability of having systematically outperformed its peer group.

Figure \ref{fig:examples2} comparies Wrigley and Tootsie Roll; Figure \ref{fig:examples3} compares Publix and Whole Foods; and Figure \ref{fig:examples4} compares Abercrombie and Finish Line Sports.  In each figure, the data have been transformed to a common $(0,1)$ scale of performance.  The block dots mark the observed $y_{ij}$'s (i.e.~the benchmarked scores from the tree model), while the black line indicates the estimated mean Gaussian-process trajectory, conditional upon that trajectory being nonzero.  We have also included draws from the posterior predictive distribution of future benchmarked ROA for the 5 years after the end of the database (2009--2013).  The posterior predictive mean is the blue line, while the shaded grey areas indicate $50\%$, $75\%$, and $95\%$ posterior-predictive credible intervals.  (The blue line and the black line will not match up in general, since the observed performance in the final year includes a contribution from both the mean trajectory and an unobserved residual.)

These figures highlight some general features of the methodology.  For one thing, it is apparent that the estimated mean trajectories are shrunk back toward the global average of zero (i.e.~the 50th percentile within a firm's peer group).  This helps regularize the estimates, since extreme outcomes are likely to be attributed to random noise even among those firms whose performance overall is more consistent with the alternative model.  Second, the model-averaged predictions of the future benchmarked scores reflect, in an intuitive way, the firm's past performance.  For example, Whole Foods is much more volatile than Publix, which translates into a larger value of $v_i$ (the residual variance) and greater uncertainty about the future.

\begin{figure}
\begin{center}
\includegraphics[width=5.5in]{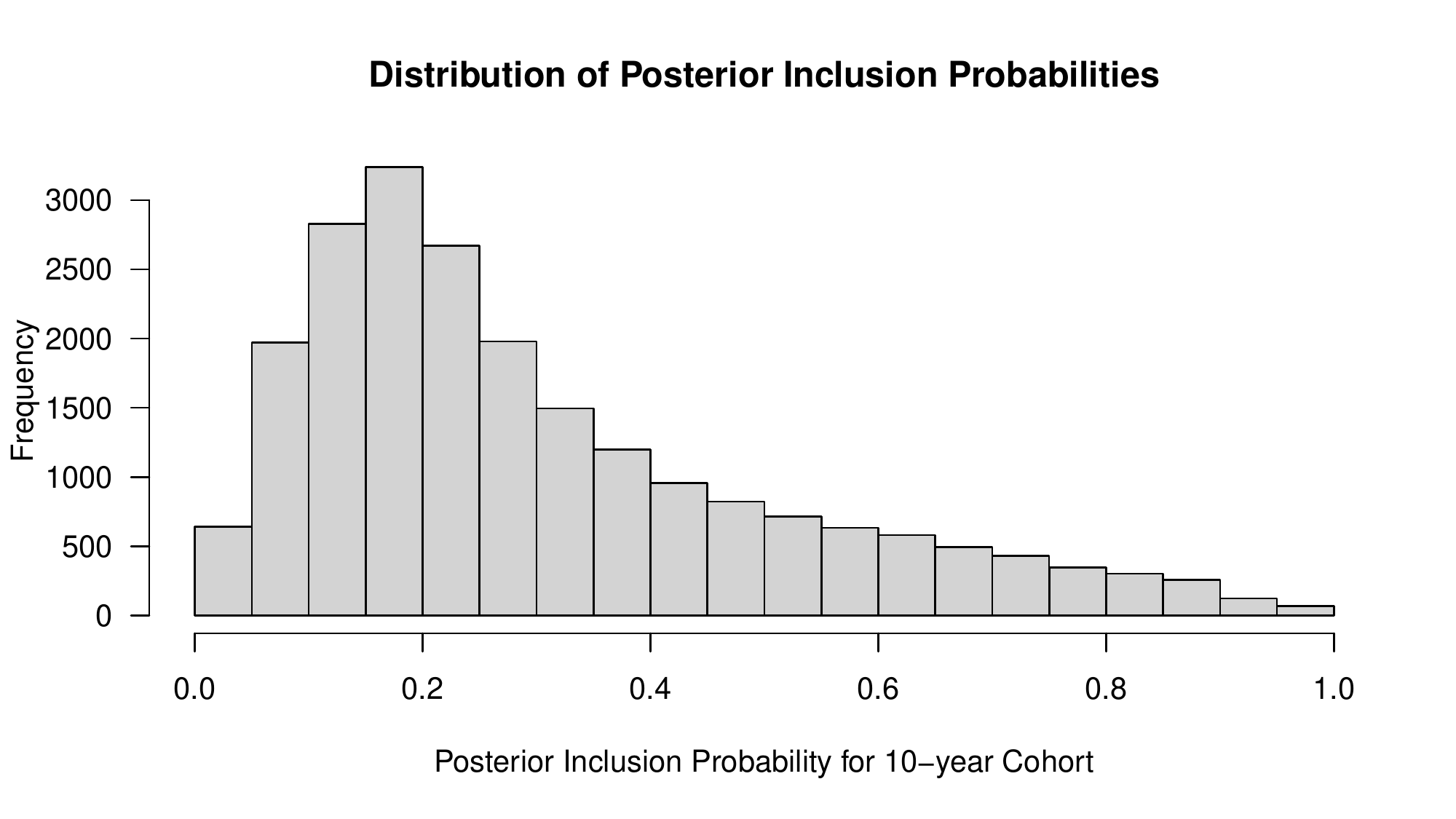}
\caption{\label{fig:postinchist}Posterior inclusion probabilities for the worldwide firms for which at least 10 years of data were available.}
\end{center}
\end{figure}

Finally, a histogram of the posterior inclusion probabilities is shown in Figure \ref{fig:postinchist}.  This histogram restricts attention to the 21,759 firms for which at least 10 years of data are available, since conclusions are necessarily vague for firms with less than 10 years of history.	Of these firms, 3,956 had inclusion probabilities greater than $50\%$, while only $193$ had posterior inclusion probabilities greater than $90\%$.  It would appear that systematically superior performance is quite rare, and that a large part of the variation in firms' performance from year to year is a consequence of short-term effects (modeled here as residuals).  This general result about the rarity of corporate outperformance is supported by other recent studies on the subject \citep[e.g.][]{henderson:etal:2009}.

\singlespace
\bibliographystyle{abbrvnat}
\bibliography{masterbib,business_articles}

\end{document}